\documentclass[11pt,a4paper]{article}
\usepackage{epsf,epsfig,amsfonts,amsgen,amsmath,amstext,amsbsy,amsopn,amsthm}
\usepackage{amssymb}
\usepackage{cite}
\usepackage{enumitem}
\theoremstyle{definition}

\begin{document}
\title{\bf {\Large Entanglement-assisted quantum MDS codes from constacyclic codes with large minimum distance}}
\date{}

\author{Liangdong Lu$^{a,b,\dag}$, Wenping Ma$^{a}$,  Ruihu Li$^{b}$,   Yuena Ma$^{b}$, Yang Liu$^{b}$, Hao Cao$^{a}$\\
 a.  State Key Laboratory of Integrated Services Networks,  Xidian University,  \\Xi'an, Shaanxi, 710051,\ China, (email:
$^{\dag}$ kelinglv@163.com, )\\
b. College of  Science,  Air Force Engineering University,  Xi'an,
 Shaanxi,\\ 710051, China}

\maketitle

\begin{abstract} The entanglement-assisted (EA) formalism allows arbitrary classical linear
codes to transform into entanglement-assisted quantum error
correcting codes (EAQECCs) by using pre-shared entanglement between
the sender and the receiver. In this work, we propose a
decomposition of the defining set of constacyclic codes. Using this
method, we  construct four classes of $q$-ary entanglement-assisted
quantum MDS (EAQMDS) codes based on classical constacyclic MDS codes
by exploiting less pre-shared maximally entangled states. We show
that a class of $q$-ary EAQMDS have minimum distance upper limit
greater than $3q-1$. Some of them have much larger minimum distance
than the known quantum MDS (QMDS) codes of the same length. Most of
these $q$-ary EAQMDS codes are new in the sense that their
parameters are not covered  by the codes available in the
literature.

\medskip

\end{abstract}

\section{\label{sec:level1} Introduction\protect}

Quantum error-correcting codes play an important role in quantum
information processing and quantum computation\cite{shor,cal1,cal2,Ketkar}.
The stabilizer formalism allows standard quantum codes to be
constructed from dual-containing (or self-orthogonal) classical
codes \cite{cal2}. However,  the dual-containing condition  forms  a barrier
in the development of quantum coding theory.
In Ref. \cite{bru06}, Brun {\it et al.} proposed
a entanglement-assisted (EA) stabilizer formalism
which shown that
non-dual-containing classical codes can be used to
construct EAQECCs  if shared
entanglement is available between the sender and receiver.

Let $q$ be a prime power. A $q$-ary  $[[n,k,d;c]]$ EAQECC  that
encodes $k$ information qubits into $n$ channel qubits with the help
of  $c$ pairs of maximally-entangled Bell states (ebits) can correct
up to $\lfloor\frac{d-1}{2}\rfloor$  errors, where  $d$ is the
minimum distance of the code. If $c=0$, then it is called a $q$-ary
standard $[[n,k,d]]$  quantum code $\mathcal{Q}$. We denote a
$q$-ary $[[n,k,d;c]]$ EAQECC by $[[n,k,d;c]]_{q}$.
Currently, many works have
focused on the construction of EAQECCs based on classical
linear codes, see
\cite{Hsi, Wil1, lai, lai2, lai3, Hsieh, Fujiwara, Wilde, Lu1, Lu2, guo}.
As in classical coding theory, one of the central tasks in quantum
coding theory is to construct quantum codes and EA-quantum
codes with the best possible minimum distance.

{\bf  Theory 1.1 \cite{bru06,lai}.} ( EA-Quantum Singleton Bound) An
$[[n,k,d;c]]_{q}$ EAQECC satisfies $$n+c-k\geq 2(d-1),$$ where
$0\leq c \leq n-1$.

 A EAQECC achieving this bound is called a EA-quantum
 maximum-distance-separable (EAQMDS) code. If $c=0$, then this bound
 is Quantum Singleton Bound, and a code achieving the bound is called
 quantum maximum-distance-separable (QMDS) code.
Just  as in the classical linear codes, QMDS codes  and EAQMDS codes
form an important family of quantum codes. Constructing QMDS codes
and EAQMDS codes had become a central topic for quantum
error-correction codes.
Many classes of QMDS codes
have been constructed by  different methods, in particular, the
constructions of obtained from constacyclic codes or negacyclic
codes containing their Hermitian dual over $F_{q^{2}}$
\cite{Li3,Chen,He,Jin,Kai1,Kai2,Wang,ZhangT1,ZhangT2}.
According to the MDS conjecture in \cite{Ketkar}, the
maximal-distance-separable (MDS) code cannot exceed $q^{2}+1$.
Therefore, for larger distance  than $q+1$ of
code length $n\leq q^2+1$, one need to construct a EA-quantum MDS code.
The following Proposition is one of the most frequently used
construction methods.

 {\bf Proposition 1.2} \cite{bru06,Wil1}.\ \   If  $\mathcal {C}$$=[n,k,d]_{q^{2}}$
is a classical code over $F_{q^{2}}$ and $H$ is its parity check
matrix, then $\mathcal{C}$$^{\perp _{h}}$ EA stabilizes an $[[n,2k-n+c,d;c]]_{q}$ EAQECC,
 where $c=$rank$(HH^{\dagger})$ is the number of maximally entangled states
required and $H^{\dagger}$ is the conjugate
 matrix of $H$  over $F_{q^{2}}$.

In resent years, many scholars have constructed
several entanglement-assisted quantum codes with good parameters in \cite{bru06,Wil1}.

 Li et al. \cite{Li1,Li2} proposed
 the concept about a decomposition of the defining set of cyclic codes,
 and they construct some good entanglement-assisted
quantum codes with the help of this concept\cite{Li1}.
Lu and Li \cite{Lu1}constructed some families of
entanglement-assisted quantum codes from primitive quaternary BCH codes.

In this paper, we construct several families of
EA-quantum MDS codes with length $n$ from classical
constacyclic codes. From \cite{Chen}, Chen
points that dual-containing constacylic codes over $F_{q^{2}}$ exist
only when the order $r$ is a divisor of $q+1$.
We construct EA-quantum MDS codes from
constacyclic codes with $r=q+1$ or negacyclic codes with $r=2$, respectively.
 More precisely, Our main contribution on new
$q$-ary quantum MDS codes is as follows:

(1) $[[q^2+1,q^2-2d+7,d;4]]_{q}$, where $q$ be an odd prime power,
$q+3\leq d \leq 3q-1$ is even.

(2) $[[\frac{q^2+1}{10},\frac{q^2+1}{10}-2d+3,d;1]]_{q}$, where $q$
be an odd prime power of the form $10m+3$, $2\leq d \leq 6m+2$ is
even.

(3) $[[\frac{q^2+1}{10},\frac{q^2+1}{10}-2d+3,d;1]]_{q}$, where $q$
be an odd prime power of the form $10m+7$, $2\leq d \leq 6m+4$ is
even.

(4) $[[\frac{q^2-1}{h},\frac{q^2-1}{h}-2d+3,d;1]]_{q}$, where $q$ be
an odd prime power, $h\in \{3,5,7\}$ is a divisor of $q+1$ and
$\frac{q+1}{h}+1\leq d \leq \frac{(q+1)(h+3)}{2h}-1$.

In construction (1), we obtain some EA-quantum MDS codes with the
minimal distance upper limit greater than $3q+2$ by consuming four
pre-shared maximally entangled states. In construction (2)-(3),
consumed only one pair of maximally entangled states,  each
EA-quantum MDS code has half length comparing with  the standard
QMDS code of the same minimum distance constructed. In construction (4),
consuming only one pair of maximally entangled states, we obtain a
family of EA-quantum MDS codes with the minimal distance larger than
the standard quantum MDS codes in Ref.\cite{Chen}. Comparing the
parameters with all known EA-quantum MDS codes, we find that these
quantum MDS  codes are new in the sense that their parameters are
not covered by the codes available in the literature.

The paper is organized as follows. In Section 2, basic notations and
results about EA-quantum codes and constacyclic codes are provided.
The concept of a decomposition of the defining set of constacyclic codes is stated,
which is extended from the decomposition of the defining set of cyclic codes
proposed by Li\cite{Li1}.
In Section 3, we give some new classes of EA-quantum MDS
codes. The conclusion is given in Section 4.

\section{\bf Preliminaries}
In this section, we review some basic results on constacyclic codes,
BCH codes, and EAQECCs for the purpose of this paper. For details on BCH codes and constacyclic
codes can be found in standard textbook on coding theory \cite{Macwilliams,Huffman}, and
 for EAQECCs please see Refs.\cite{bru06,Hsi, Wil1, lai, lai2, lai3, Hsieh, Fujiwara, Wilde, Lu1, Lu2}.

Let $p$ be a prime number and $q$ a power of $p$, ie., $q=p^{l}$ for
some $l>0$. $F_{q^{2}}$ denotes the finite field with $q^{2}$
elements. For any $\alpha \in F_{q^{2}}$, the conjugation of
$\alpha$ is denoted by $\overline{\alpha}=\alpha^{q}$. Given two
vectors $\mathbf{x}=(x_{1},x_{2},\cdots,x_{n})$ and
$\mathbf{y}=(y_{1},y_{2},\cdots,y_{n})\in F_{q^{2}}^{n}$, their
Hermitian inner product is defined as
$(\mathbf{x},\mathbf{y})_{h}=\sum \overline{x_{i}}y_{i}=\overline{x_{1}}y_{1}+\overline{x_{2}}y_{2}+\cdots+\overline{x_{n}}y_{n}.$
For a linear code $\mathcal{C}$ over $F_{q^{2}}$ of length $n$, the
Hermitian dual code $\mathcal{C}^{\bot _{h}}$ is defined as
 $\mathcal{C}^{\bot _{h}}=\{x\in  F_{q^{2}}^{n} | (x,y)_{h}=0, \forall  y $$\in \mathcal{C}\}$.
If $\mathcal{C} \subseteq \mathcal{C}^{\bot _{h}}$, then
$\mathcal{C}$ is called a Hermitian dual containing code, and
$\mathcal{C}^{\bot _{h}}$ is called a Hermitian self-orthogonal
code.

We now recall some results about classical constacyclic codes,
negacyclic code and cyclic codes. For any
vector $(c_{0},c_{1},\cdots,c_{n-1}) $
$\in F_{q^{2}}^{n}$, a
$q^{2}$-ary linear code $\mathcal{C}$ of length $n$ is called $\eta$-constacyclic
if it is invariant
under the $\eta$-constacyclic shift of $F_{q^{2}}^{n}$:
$$
(c_{0}, c_{1}, \cdots , c_{n-1}) \rightarrow (\eta c_{n-1},
c_{0},\cdots, c_{n-2}),$$ where $\eta$ is a nonzero element of
$F_{q^{2}}$. Moreover, if
$\eta=1$, then $\mathcal{C}$ is called a cyclic code ; and if $\eta=-1$,
 then $\mathcal{C}$ is  called a negacyclic code.

 For a constacyclic code $\mathcal{C}$, each codeword $c =
(c_{0}, c_{1}, \cdots, c_{n-1})$ is customarily represented in its
polynomial form: $c(x) = c_{0} + c_{1}x + \cdots + c_{n-1}x_{n-1},$
and the code $\mathcal{C}$ is in turn identified with the set of all
polynomial representations of its codewords. The proper context for
studying  constacyclic codes is the residue class ring
$\mathcal{R}_{n}=\mathbb{F}_{q}[x]/(x^{n}-\eta)$. $xc(x)$ corresponds
to a constacyclic shift of $c(x)$ in the ring $\mathcal{R}_{n}$. As we
all know, a linear code $\mathcal{C}$ of length $n$ over $F_{q^{2}}$
is constacyclic if and only if C is an ideal of the quotient ring
$\mathcal{R}_{n}=\mathbb{F}_{q}[x]/(x^{n}-\eta)$. It follows that
$\mathcal{C}$ is generated by monic factors of $(x^{n}-\eta)$, i.e.,
$\mathcal{C}=\langle f(x) \rangle$ and $f(x)|(x^{n}-\eta)$. The $f(x)$
is called the generator polynomial of $\mathcal{C}_{n}$.

Let $\eta \in F_{q^{2}}$ be a primitive $r$th root of unity. Let
 $gcd(n,q)=1$, then there exists a primitive $rn$-th root
of unity $\omega$ in some extension field field of $F_{q^{2}}$ such that
$\omega^{n}=\eta$. Hence, $x^{n}-\eta =\prod ^{n-1}
_{i=0} (x-\omega^{1+ir})$.
Let $\Omega=\{1+ir|0\leq i \leq n-1\}$. For each $j\in \Omega$, let
$C_{j}$ be the $q^{2}$-cyclotomic coset modulo $rn$ containing $j$.
Let $\mathcal{C}$ be an $\eta$-constacyclic code of length $n$ over $F_{q^{2}}$
with generator polynomial $g(x)$. The set  $T=\{j\in\Omega|g(\omega^{j})=0\}$ is
called the defining set of  $\mathcal{C}$.
Let $s$ be an integer with $0\leq s < rn$, the
$q^{2}$-cyclotomic coset modulo $rn$ that contains $s$ is defined by
the set $C_{s}=\{s, sq^{2}, sq^{2\cdot 2}, \cdots, sq^{2(k-1)} \}$
(mod $rn$), where $k$ is the smallest positive integer such that
$xq^{2k}$ $\equiv x$ (mod $rn$).
We can see that the defining set $T$ is a union of some
$q^{2}$-cyclotomic cosets module $rn$ and $dim(\mathcal{C}) =
n-|T|$.

{\bf Lemma 2.1 \cite{Kai2}}. Let $\mathcal{C}$ be a $q^{2}$-ary
constacyclic code of length $n$ with defining set $T$. Then
$\mathcal{C}$ contains its Hermitian dual code if and only if $T
\bigcap T^{-q}=\emptyset$, where $T^{-q}$ denotes the set
$T^{-q}=\{-qz($mod  $rn)| z\in T\}$.

Let $\mathcal {C}$ be a constacyclic code with a defining set $T =
\bigcup \limits_{s \in S} C_{s}$. Denoting $T^{-q}=\{rn-qs | s\in T
\}$, then we can deduce that the  defining set of $\mathcal
{C}$$^{\bot _{h}}$ is $T^{\perp _{h}} =$$ \mathbb{Z}_{n}
$$\backslash T^{-q}$, see Ref. \cite{Kai2}.

Since there is a striking similarity between cyclic codes and
constacyclic code, we give a correspondence defining of skew aymmetric
and skew asymmetric as follows.

 A cyclotomic coset $C_{s}$ is {\it skew symmetric } if $rn-qs$ mod $rn\in
 C_{s}$; and otherwise is skew asymmetric otherwise. {\it  Skew asymmetric
 cosets}
$C_{s}$ and $C_{rn-qs}$ come in pair, we use $(C_{s},C_{rn-qs})$ to
denote such a pair.

The following results on $q^{2}$-cyclotomic  cosets, dual containing
constacyclic codes are bases of our discussion.

{\bf Lemma 2.2 \cite{Li2,Kai2}.} Let $r$ be a positive divisor of
$q+1$ and $\eta\in F^{*}_{q^{2}}$ be of order $r$. Let $\mathcal{C}$
be a $\eta$-constacyclic code of length $n$ over $F_{q^{2}}$ with
defining set $T$, then $\mathcal {C}$$^{\perp
_{h}}$$\subseteq\mathcal{C}$ if and only if one of the following
holds:

 (1) $T \cap$$T^{-q}=\emptyset$, where $T^{-q}=\{rn-qs \mid s\in
T\}$.

 (2) If $i,j,k\in T$, then
 $C_{i}$ is not a skew asymmetric coset and
($C_{j}$, $C_{k}$) is not  a skew asymmetric cosets pair.

According to Lemma 2.2, $\mathcal{C}^{\perp _{h}}\subseteq$
$\mathcal{C}$ can be described by the relationship of its cyclotomic
coset $C_{s}$. However, a defining set $T$ of a non-dual-containing
(or non-self-orthogonal) classical codes is $T \cap$$T^{-q}\neq
\emptyset$. We gave a definition for decomposition of the defining
set of cyclic codes in \cite{Li1,Lu1} and Chen et al. defined a
decomposition of the defining set of negacyclic codes in
\cite{Chen2}. In order to construct  EA-quantum MDS codes for larger
distance than $q+1$ of code length $n\leq q^2+1$, we introduce a
fundamental definition of decomposition of the defining set of
constacyclic codes.

{\bf Definition 2.1}  {\it  Let $\eta \in F_{q^{2}}$ be a primitive
$r$th root of unity. Let $ \mathcal {C}$ be a $\eta$-constacyclic
code of length $n$ with defining set $T$. Denote $T_{ss}=T
\cap$$T^{-q}$ and $T_{sas}=T \setminus
$$T_{ss}$, where $T^{-q}=\{rn-qx | x\in T \}$ and $r$ is a factor of $q+1$. $T=T_{ss} \cup
T_{sas}$ is called decomposition of the defining set of
$\mathcal{C}$.}

To  determine $T_{ss}$ and $T_{sas}$, we give the following lemma to
characterize them.

{\bf Lemma 2.3 \cite{Li2,Kai2}.} Let $r$ be a positive divisor of
$q+1$  and $\eta\in F^{*}_{q^{2}}$ be of order $r$. Let $gcd(q, n) =
1$, $ord_{rn}$$(q^{2})=m$, $0 \leq x, y$, $z \leq n-1$.

(1) $C_{x}$ is skew symmetric if and only if there is a $t\leq
\lfloor\frac{m}{2}\rfloor$
 such that $x \equiv xq^{2t+1}$(mod rn).

(2) If $C_{y}\neq C_{z}$, $(C_{y}, C_{z})$ form a skew asymmetric
pair if and only if there is a $t\leq \lfloor\frac{m}{2}\rfloor$
such that $y \equiv zq^{2t+1}$ (mod n) or $z \equiv yq^{2t+1}$(mod
rn).

Using the decomposition of a  defining set  $T$ of a constacyclic
code $\mathcal{C}$,  one  can give a decomposition of
$\mathcal{C}$$^{\perp_{h}}$ as follow.

{\bf Lemma 2.4}  Let $\mathcal {C}$  be a constacyclic code with
defining set $T$, $T=T_{ss} \cup T_{sas}$ be decomposition of $T$.
 Denote the constacyclic codes $\mathcal{C}$$_{R}$ and  $\mathcal{C}$$_{E}$ with defining set
 $T_{sas}$ and $T_{ss}$, respectively.
 Then $\mathcal{C}$$^{\perp _{h}}_{R}$$\subseteq \mathcal{C}$$_{R}$,
 $\mathcal{C}$$_{E}$$\cap$ $\mathcal{C}$$^{\perp_{h}}_{E}$ $=\{0\}$,
$\mathcal{C}$$^{\perp_{h}}_{R}$ $\subset \mathcal{C}$$_{E}$,
$\mathcal{C}$$_{R}$$\cap$ $\mathcal{C}$$_{E}$ $=\mathcal{C}$ and
 $\mathcal{C}$$^{\perp_{h}}_{R}$ $+$ $\mathcal{C}$$^{\perp_{h}}_{E}$ $=\mathcal{C}$$^{\perp_{h}}$.

Using the decomposition of a  defining set  $T$, one can calculate
the number of needed ebits with a algebra method.

 {\bf Lemma 2.5.} Let $T$ be a defining set of a constacyclic
code $ \mathcal {C}$, $T=T_{ss}\cup T_{sas}$ be decomposition of
$T$. Using $\mathcal{C}$$^{\perp_{h}}$ as EA stabilizer, the optimal
number of needed  ebits is $c=\mid T_{ss} \mid$.

  {\bf   Proof.}\ \ According to Definition 2.1, we denote
  the defining sets of constacyclic codes $\mathcal{C}_{1}$ and
  $\mathcal{C}_{2}$ into $T_{ss}$ and $T_{sas}$, respectively. The
  parity check matrix of $\mathcal{C}_{1}$ and
  $\mathcal{C}_{2}$ are $H_{1}$ and $H_{2}$, respectively. Let  $H=\left(
\begin{array}{lllllll}
H_{1}\\
H_{2}\\
\end{array} \right)$ be the parity check matrix of $\mathcal{C}$.
Then, $$HH^{\dag}=\left(
\begin{array}{lllllll}
H_{1}H_{1}^{\dag} & H_{1}H_{2}^{\dag}\\
H_{2}H_{1}^{\dag} &  H_{2}H_{2}^{\dag}\\
\end{array} \right).$$

Since $H_{2}$ is the parity check matrix of $\mathcal{C}_{2}$ with
defining set of $T_{sas}$,  $H_{2}H_{2}^{\dag}=0$. Because of
$\mathcal{C}_{1}^{\bot h} \subseteq \mathcal{C}_{2}$,
$H_{1}H_{2}^{\dag}=0$ and  $H_{2}H_{1}^{\dag}=0$. Therefore,

$$HH^{\dag}=\left(
\begin{array}{lllllll}
H_{1}H_{1}^{\dag} & 0\\
0 &  0\\
\end{array} \right).$$ According to \cite{bru06,Hsi,Wil1}, one
obtain that $c=rank(HH^{\dag})=rank(H_{1}H_{1}^{\dag})$. Since
$H_{1}$ is the parity check matrix of $\mathcal{C}_{2}$ with
defining set of $T_{ss}$, $H_{1}$ is a full-rank matrix. Hence,
$c=rank(H_{1}H_{1}^{\dag})=|T_{ss}|$.

{\bf Lemma 2.6 \cite{Yang}.} (The BCH bound for Constacyclic Codes)
Let $\mathcal{C}$ be an $\eta$-constacyclic code of length $n$ over
 $F_{q^{2}}$, where $\eta$ is a primitive $r$th root of unity. Let $\omega$
 be a primitive $rn$-th root of unity in an extension field of $F_{q^{2}}$ such that
 $\omega^{n}=\eta$. Assume the generator polynomial of $\mathcal{C}$ has roots that
 include the set $\{\omega^{1+ri}|i_{1}\leq i \leq i_{1}+d-2\}$. Then the minimum distance of $\mathcal{C}$ is at least $d$.

{\bf Theorem 2.7.} Let $\mathcal{C}$ be an $[n,k,d]_{q^{2}}$
Constacyclic code with defining set $T$,  and the  decomposition of
$T$ be $T=T_{ss}\cup T_{sas}$. Then $\mathcal{C}$$^{\perp_{h}}$  EA
stabilizes an $q$-ary $[[n,n-2|T|+|T_{ss}|,d \geq \delta ; |T_{ss}|]]$
EAQECC.

  {\bf   Proof.}\ \ The dimension of $\mathcal{C}$ is $k=n-|T|$. From Proposition 1.2, Lemma 2.5 and
 Lemma 2.6, we know $\mathcal{C}$$^{\perp_{h}}$  EA stabilizes an $q$-ary
EAQECC with parameters $[[n,2k-n+c,d ;c]]$ $=[[n,n-2|T|+|T_{ss}|,d;
|T_{ss}|]]$.

\section{New  EA-quantum MDS Codes}

In this section, we consider $\eta$-constacyclic codes over
$F_{q^{2}}$ of length $n$ to construct EA-quantum codes. To do this,
we give a sufficient condition for a decomposition of the defining
set of $\eta$-constacyclic codes over $F_{q^{2}}$ of length $n$
which do not contain their Hermitian duals. First, we compute
$q^{2}$-cyclotomic cosets modulo $rn$ where $r=q+1$(constacyclic
codes) or $r=2$(negacyclic codes).

\subsection{EA-quantum MDS Codes of Lenght
$n=q^{2}+1$ with $q\equiv 1$ mod $4$}

In this subsection, we construct some classes of $q$-ary  EA-quantum
MDS Codes of Length $n=q^{2}+1$, where $q$ be a prime or an odd
prime power of the form $q\equiv 1$ mod $4$. First, Let $r=2$, we
consider negacyclic codes over $F_{q^{2}}$ of length $n$ to
construct EA-quantum codes. To do this, we give a  decomposition of
defining set sufficient condition for negacyclic codes over
$F_{q^{2}}$ of length $n$ which do not contain their Hermitian
duals.

 {\bf  Lemma 3.1 \cite{Kai1}:}  Let $n=q^{2}+1$ with $q\equiv 1$ mod $4$, and $s=\frac{n}{2}$.
 Then the $q^{2}-$cyclotomic cosets modulo $2n$ containing odd integers form 1 to $2n$ are
 $C_{s}=\{s\}$ and $C_{s+2i}=\{s+2i,s-2i\}$ for $1\leq i\leq s-1$.

 {\bf  Lemma 3.2:}   Let $n=q^{2}+1$ with $q\equiv1$ mod $4$, $q\geq 5$ and $s=\frac{n}{2}$.
 If $\mathcal{C}$ is a $q^{2}$-ary negacyclic code of length
$n$ with define set $T=\bigcup_{i=0}^{k}C_{s+2i}$, where $0\leq
k\leq
 \frac{3q-3}{2}$, and the decomposition of a  defining set
 $T=T_{ss}\bigcup T_{sas}$, then $(C_{s+q+1},C_{s-q+1})$ forms a skew-asymmetric pair. And
$$|T_{ss}| = \left\{
\begin{array}{lll}
0,                 &\mbox {if $0  \leq k \leq \frac{q-1}{2}$;}\\
4,               &\mbox {if $\frac{q+1}{2} \leq k \leq \frac{3q-3}{2}$.}\\
 \end{array}
\right. $$

 {\bf  Proof:} (i) Since $q\equiv1$ mod $4$, $4|(q-1)$, then
 $-(s+q+1)q\equiv
 -[\frac{q-1}{4}\cdot 2(q^2+1)+\frac{1}{4}\cdot2(q^2+1)+q^2+q]$ $\equiv 2n-(\frac{q^2+1}{2}+q^2+q)$
 $\equiv 2q^2+2-(\frac{q^2+1}{2}+q^2+q)$  $\equiv
 \frac{q^2+1}{2}-q+1$ $\equiv s-q+1$ mod $2n$. Hence, $(C_{s+q+1},C_{s-q+1})$
 forms
 a skew-asymmetric pair.

(ii)Let $T=\bigcup_{i=0}^{k}C_{s+2i}$, where $0\leq k\leq
 \frac{3q-3}{2}$.
According to Ref.\cite{Kai1}, one can obtain that for $0\leq k\leq
 \frac{q-1}{2}$,  $\mathcal{C}^{\perp h}\subseteq \mathcal{C}$. It
 means that $|T_{ss}| =0$ if $0\leq k\leq
 \frac{q-1}{2}$.

For $\frac{q+1}{2}\leq k\leq
 \frac{3q-3}{2}$,  since $(C_{s+q+1},C_{s-q+1})$
 forms
 a skew-asymmetric pair, $T_{ss}$ comprises the set $\{C_{s+q+1},C_{s-q+1}\}$ at least.
 According to the concept about a  decomposition of the  defining set $T$,
 one obtain that $T_{sas}=T\backslash T_{ss}$. In order to testify $|T_{ss}| =4$ if
$\frac{q+1}{2}\leq k\leq \frac{3q-1}{2}$, from Definition 2.1 and
Lemma 2.2,
 we need  to testify that there is no skew symmetric cyclotomic
 coset, and any two cyclotomic
 coset do not form a  skew asymmetric pair in  $T_{sas}$.

Let $I=\{s+2i|0\leq i\leq \frac{3q-3}{2}\}$,   $I_{1}=\{s+2i|0\leq
i\leq \frac{q-1}{2}\}$, $I_{2}=\{s+2i|\frac{q+3}{2}\leq i\leq
\frac{3q-3}{2}\}$. Then $I=I_{1}\cup I_{2} \cup \{s+q+1\}$. Only we
need  to testy that for $\forall x\in I_{1}\cup I_{2}$, then $-qx$
(mod $2n$)$\not \in I_{1}\cup I_{2}$.

Let $x,y\in I_{1}\cup I_{2}$. From  Lemma 2.3,  $C_{x}$ is not a
skew symmetric cyclotomic  coset, and any $C_{x},C_{y}$ do not form
a skew asymmetric pair if and only if $x+yq\not\equiv0$ mod $2n$.

For $ x,y\in I_{1}$, then $\frac{q-1}{4}(2n)
<\frac{q-1}{4}(2n)+q^2+1=s(q+1)\leq x+yq \leq
(s+q-1)(q+1)=s(q+1)+(q-1)(q+1)=\frac{q-1}{4}(2n)+2q^2<(\frac{q-1}{4}+1)(2n)$.

For $ x,y\in I_{2}$, then $(\frac{q-1}{4}+1)(2n)
<\frac{q-1}{4}(2n)+q^2+1+(q^2+1)+4(q+1)=(s+q+3)(q+1)\leq x+yq \leq
(s+3q-3)(q+1)=s(q+1)+(3q-3)(q+1)=\frac{q-1}{4}(2n)+4(q^2+1)-6<(\frac{q-1}{4}+2)(2n)$.

For $ x\in I_{1} \cup I_{2}$ and $y\in I_{2}$, then
$(\frac{q-1}{4}+1)(2n)
<\frac{q-1}{4}(2n)+2(q^2+1)+3q-1=s+(s+q+3)(q+1)\leq x+yq \leq
(s+3q-3)(q+1)=s(q+1)+(3q-3)(q+1)=\frac{q-1}{4}(2n)+4(q^2+1)-6<(\frac{q-1}{4}+2)(2n)$.

Hence, there is no skew symmetric cyclotomic  cosets, and any two
cyclotomic  coset do not form a skew asymmetric pair. That implies
that $T_{ss}=\{C_{s+q+1},C_{s-q+1}\}=\{s+q+1,s-q-1,s-q+1,s+q-1\}$
and $|T_{ss}|=4$  for $\frac{q+1}{2} \leq k \leq \frac{3q-3}{2}$,
when the defining set $T=\bigcup_{i=0}^{k}C_{s+2i}$, where $0\leq
k\leq
 \frac{3q-3}{2}$.

{\bf  Theory 3.3:}   Let  $q$ be an odd prime power and $q\equiv1$
mod $4$. Then there exists a q-ary $[[q^2+1,q^2-2d+7,d;4]]$-
EA-quantum MDS codes, where $q+3\leq d \leq 3q-1$ is even.

 {\bf  Proof:} Consider the negacyclic codes over $F_{q^{2}}$ of
length $n=q^2+1$ with defining set $T=\bigcup_{i=0}^{k}C_{s+2i}$,
where $0 \leq k \leq \frac{3q-3}{2}$ for $q$ be an odd prime power
and $q \equiv 1$ mod 4. By Lemma 3.2 (i), there is $c=|T_{ss}|=4$ if
$\frac{q+1}{2} \leq k \leq \frac{3q-3}{2}$. Since every
$q^{2}$-cyclotomic coset $C_{x}=\{x,n-x\}$ and $x$ must be odd
number, we can obtain that $T$ consists of $2k+1$ integers
$\{n-(s+2k),\cdots,n-s-2,s,s+2,\cdots,s+2k\}$. It implies that
$\mathcal{C}$ has minimum distance at least $2k+2$. Hence,
$\mathcal{C}$ is a $q^{2}$-ary negacyclic code with parameters
$[n,n-2(k+1)+1,\geq 2k+2]$. Combining Theory 2.7 with EA-quantum
Singleton bound, we can obtain a EA-quantum MDS code with parameters
$[[q^2+1,q^2-2d+7,d;4]]_{q}$, where $q+3\leq d \leq 3q-1$ is even.

\begin{center}
Table 1 EA-quantum MDS codes with length $n=q^{2}+1$ for $q\equiv1$ mod 4\\
\begin{tabular}{lllllllllll}
  \hline
  q               &n      &$[[n,k,d;4]]_{q}$           & d                    &d   in \cite{Chen2} \\
\hline
 9                &82     &$[[82,88-2d,d;4]]_{9}$       & $12\leq d\leq26$ is even     & $14\leq d\leq18$ is even\\

  13               &170    &$[[170,176-2d,d;4]]_{13}$    & $16\leq d\leq38$ is even     & $18\leq d\leq26$ is even\\

 17               &290    &$[[290,296-2d,d;4]]_{17}$    & $20\leq d\leq50$ is even     & $22\leq d\leq34$ is even \\

 19               &362    &$[[362,368-2d,d;4]]_{19}$    & $22\leq d\leq56$ is even     & $24\leq d\leq38$ is even \\

 25               &626    &$[[626,632-2d,d;4]]_{25}$    & $28\leq d\leq74$ is even     & $30\leq d\leq50$ is even\\

 29               &842    &$[[842,848-2d,d;4]]_{29}$    & $32\leq d\leq86$ is even     & $34\leq d\leq58$ is even\\

 \hline
  \end{tabular}
\end{center}

\subsection{ EA-quantum MDS Codes of Lenght
$n=q^{2}+1$ with $q\equiv 3$ mod $4$}

In this subsection, we construct some classes of $q$-ary  EA-quantum
MDS Codes of Length $n=q^{2}+1$, where $q$ be a prime or an odd
prime power of the form $q\equiv 3$ mod $4$. Let $r=q+1$ and $\eta\in
F_{q^{2}}$ be a primitive $r$th root of unity.
To do this, we give a decomposition of
defining set sufficient condition for $\eta$-constacyclic codes over
$F_{q^{2}}$ of length $n$ which do not contain their Hermitian
duals.

 {\bf  Lemma 3.4 \cite{Kai2}:}  Let $n=q^{2}+1$ with $q\equiv 3$ mod $4$, and $s=\frac{n}{2}$.
 Then the $q^{2}-$cyclotomic cosets modulo $(q+1)n$ containing odd integers form 1 to $(q+1)n$ are
 $C_{s}=\{s\}$ and $C_{s+(q+1)i}=\{s+(q+1)i,s-(q+1)i\}$ for $1\leq i\leq s-1$.

 {\bf  Lemma 3.5:}   Let $n=q^{2}+1$ with $q\equiv 3$ mod $4$, $q\geq 7$ and
 $s=\frac{n}{2}$. If $\mathcal{C}$ is a $q^{2}$-ary constacyclic code of length
$n$ with define set $T=\bigcup_{i=0}^{k}C_{s+(q+1)i}$, where $0\leq
k\leq \frac{3q-3}{2}$, and the decomposition of a  defining set
 $T=T_{ss}\bigcup T_{sas}$, then
 $(C_{s+\frac{(q+1)}{2}(q+1)},C_{s+\frac{(q-1)}{2}(q+1)})$
forms a skew asymmetric pair. And
 $$|T_{ss}| = \left\{
\begin{array}{lll}
0,                 &\mbox {if $0  \leq k \leq \frac{q-1}{2}$;}\\
4,               &\mbox {if $\frac{q+1}{2} \leq k \leq \frac{3(q-1)}{2}$.}\\
 \end{array}
\right. $$

 {\bf  Proof:}  Let $r=q+1$.

 (a) From Lemma 3.4, $C_{1}=C_{s-\frac{(q-1)}{2}(q+1)}=C_{s+\frac{(q-1)}{2}(q+1)}$,
  Since $-(s+\frac{(q+1)^{2}}{2})q\equiv
 -[r(q^2+1)+r\frac{q-1}{2}-\frac{(q^2+1)}{2}]$ $\equiv -(\frac{q^2-1}{2}-\frac{q^2+1}{2})$
 $\equiv 1$ mod $rn$,
  $(C_{s+\frac{(q+1)}{2}(q+1)},C_{s+\frac{(q-1)}{2}(q+1)})$
 forms
 a skew-asymmetric pair.

 (b) Let $T=\bigcup_{i=0}^{k}C_{s+ri}$, where $0\leq k\leq
 \frac{3q-3}{2}$. Let $I=\{s+ri|0\leq i\leq \frac{3q-3}{2}\}$,   $I_{1}=\{s+ri|0\leq
i\leq \frac{q-1}{2}\}$, $I_{2}=\{s+ri|\frac{q+3}{2}\leq i\leq
\frac{3q-3}{2}\}$. Then $I=I_{1}\cup I_{2} \cup
\{C_{s+\frac{(q-1)}{2}(q+1)}\}$. Only we need  to testy that for
$\forall x\in I_{1}\cup I_{2}$, then $-qx$ (mod $rn$)$\not \in
I_{1}\cup I_{2}$.

Let $x,y\in I_{1}\cup I_{2}$. From  Lemma 2.3,  $C_{x}$ is not a
skew symmetric cyclotomic  coset, and any $C_{x},C_{y}$ do not form
a skew asymmetric pair if and only if $x+yq\not\equiv0$ mod $rn$.

For $ x,y\in I_{1}$, then $ s(q+1)\leq x+yq \leq
[s+(q+1)\frac{q-1}{2}](q+1)=rq^2<rn$.  $T_{ss} =\emptyset $ and
$|T_{ss}|=0$ if $0\leq k\leq \frac{q-1}{2}$.

For $\frac{q+1}{2} \leq k \leq \frac{3(q-1)}{2}$,  since
$(C_{s+\frac{(q+1)}{2}(q+1)},C_{s+\frac{(q-1)}{2}(q+1)})$
 forms
 a skew-asymmetric pair, $T_{ss}$ comprises the set $\{C_{s+\frac{(q+1)}{2}(q+1)},C_{s+\frac{(q-1)}{2}(q+1)}\}$ at least.
 According to the concept about a  decomposition of the  defining set $T$,
 one obtain that $T_{sas}=T\backslash T_{ss}$. In order to testify $|T_{ss}| =4$ if
$\frac{q+1}{2}\leq k\leq \frac{3q-3}{2}$, from Definition 2.1 and
Lemma 2.2,
 we need  to testify that there is no skew symmetric cyclotomic
 coset, and any two cyclotomic
 coset do not form a  skew asymmetric pair in  $T_{sas}$.

For $ x,y\in I_{2}$, then
$rn<r(q^2+1)+r(2q+1)=[s+\frac{q+3}{2}(q+1)](q+1)\leq x+yq \leq
[s+(q+1)\frac{3(q-1)}{2}](q+1)=\frac{q^2+1}{2}r+\frac{3(q-1)}{2}(q+1)r=2rq^2<2(rn)$.
For $ x\in I_{1} \cup I_{2}$ and $y\in I_{2}$, then $rn
<\frac{2q^2+3q+1}{2}r=s+[s+(q+1)\frac{q+3}{2}]q\leq x+yq \leq
[s+\frac{3(q-1)}{2}(q+1)](q+1)=2rq^2<2(rn)$.

Hence, there is no skew symmetric cyclotomic cosets, and any two
cyclotomic coset do not form a skew asymmetric pair. That implies
that
$T_{ss}=\{C_{s+\frac{(q+1)}{2}(q+1)},C_{s+\frac{(q-1)}{2}(q+1)}\}=
\{s+\frac{(q+1)}{2}(q+1),s-\frac{(q+1)}{2}(q+1),s+\frac{(q-1)}{2}(q+1),s-\frac{(q-1)}{2}(q+1)\}$
and $|T_{ss}|=4$  for $\frac{q+1}{2} \leq k \leq \frac{3(q-1)}{2}$,
when the defining set $T=\bigcup_{i=0}^{k}C_{s+ri}$, where $0\leq
k\leq
 \frac{3q-3}{2}$.\\

{\bf  Theory 3.6:}   Let  $q$ be an odd prime power and $q\equiv 3$
mod $4$. Then there exists a q-ary $[[q^2+1,q^2-2d+7,d;4]]$-
EA-quantum MDS codes, where $q+3\leq d \leq 3q-1$ is even.

 {\bf  Proof:} Consider the constacyclic codes over $F_{q^{2}}$ of length
$n=q^2+1$ with defining set $T=\bigcup_{i=0}^{k}C_{s+ri}$, where $0
\leq k \leq \frac{3q-3}{2}$ for $q$ be an odd prime power and $q
\equiv 3$ mod 4. By Lemma 3.5, there is $c=|T_{ss}|=4$ if
$\frac{q+1}{2} \leq k \leq \frac{3(q-1)}{2}$. Since every
$q^{2}$-cyclotomic coset $C_{s+x}=\{s+x,s-x\}$ and $s+x$ must be odd
number, we can obtain that $T$ consists of $2k+1$ integers
$\{s-rk,\cdots,s-r,s,s+r,\cdots,s+rk\}$ (mod $rn$). It implies that
$\mathcal{C}$ has minimum distance at least $2k+2$. Hence,
$\mathcal{C}$ is a $q^{2}$-ary constacyclic code with parameters
$[n,n-2(k+1)+1,\geq 2k+2]$. Combining Theory 2.7 with EA-quantum
Singleton bound, we can obtain a EA-quantum MDS code with parameters
$[[q^2+1,q^2-2d+7,d;4]]_{q}$, where $q+3\leq d \leq 3q-1$ is even.

\begin{center}
Table 2 EA-quantum MDS codes with length $n=q^{2}+1$ for $q\equiv3$ mod 4\\
\begin{tabular}{lllllllllll}
  \hline
  q               &n      &$[[n,k,d;4]]_{q}$           & d                    \\
\hline
 7                &50     &$[[50,56-2d,d;4]]_{7}$       & $10\leq d\leq20$ is even      \\

 11               &122    &$[[122,128-2d,d;4]]_{11}$    & $14\leq d\leq32$ is even     \\

 19               &362    &$[[362,368-2d,d;4]]_{19}$    & $22\leq d\leq56$ is even       \\

 23               &530    &$[[530,536-2d,d;4]]_{23}$    & $26\leq d\leq68$ is even      \\

 31               &962    &$[[962,968-2d,d;4]]_{31}$    & $34\leq d\leq92$ is even      \\

 43               &1850   &$[[1850,1856-2d,d;4]]_{43}$  & $46\leq d\leq128$ is even      \\

 \hline
  \end{tabular}
\end{center}

\subsection{ EA-quantum MDS Codes of Lenght $n=\frac{q^{2}+1}{10}$}

In this subsection, we construct new $q$-ary  EA-quantum MDS Codes
of Length $n=\frac{q^{2}+1}{10}$, where $q$ be an odd prime power of
the form $q=10m+3$ or $q=10m+7$. Let $r=2$ and $\eta\in F_{q^{2}}$
be a primitive $r$th root of unity. To do this, we give the
well-known result of $q^{2}-$cyclotomic coset $C_{i}$ modulo $2n$
for each $i\in Z_{2n}$.

{\bf  Lemma 3.7:}   Let $n=\frac{q^{2}+1}{10}$. Then the
$q^{2}$-cyclotomic coset $C_{i}$ modulo $2n$ containing all odd
integers form $1$ to $2n$ are $C_{n}=\{n\}$ and
$C_{2i-1}=\{2i-1,2i+1\}$ for $1\leq i \leq \frac{n-1}{2}$.

{\bf  Lemma 3.8:}   Let $q$ be and prime power of form $10m+3$  and
$n=\frac{q^{2}+1}{10}$. If $\mathcal{C}$ is a $q^{2}$-ary negacyclic
code of length $n$ with define set $T=\bigcup_{i=0}^{k}C_{n+2i}$,
where $0\leq k\leq 3m$, and the decomposition of a  defining set
 $T=T_{ss}\bigcup T_{sas}$, then

\quad(i) $C_{n}$ is  skew symmetric. \quad(ii) $|T_{ss}|=1$, if $0
\leq k \leq 3m$.

 {\bf  Proof:} Since $-nq\equiv
 -(\frac{q^2+1}{5}\cdot\frac{q-1}{2}-\frac{q^2+1}{10})$ $\equiv \frac{q^2+1}{10}$ $\equiv n$ mod $2n$,
  $C_{n}$ is  skew symmetric.

Let $T=\bigcup_{i=0}^{k}C_{n+2i}$, where $0\leq k\leq 3m$. Since
$C_{n}$ is skew symmetric, $T_{ss}$ comprises the set $\{C_{n}\}$ at
least.
 According to the concept about a  decomposition of the  defining set $T$,
 one obtain that $T_{sas}=T\backslash T_{ss}$. In order to testify $|T_{ss}| =1$ if
$0\leq k\leq 3m$, from Definition 2.1 and Lemma 2.2,
 we need  to testify that there is no skew symmetric cyclotomic
 coset, and any two cyclotomic
 coset do not form a  skew asymmetric pair in  $T_{sas}$.
Let $I=\{n+2i|1\leq i\leq 3m\}$. Only we need  to testy that for
$\forall x\in I$, $-qx$ (mod $2n$)$\not \in I$ and
$T_{ss}=\{C_{n}\}$. That implies that if $x,y\in I$, from  Lemma
2.3, $C_{x}$ is not a skew symmetric cyclotomic  coset, and any
$C_{x},C_{y}$ do not form a skew asymmetric pair if and only if
$x+yq\not\equiv0$ mod $2n$.

Divide $I$ into three parts $I_{1}=[n+2,n+2m]$,
$I_{2}=[n+2m+2,n+4m]$ and $I_{3}=[n+4m+2,n+6m]$. $q=10m+3$ and
$n=\frac{q^2+1}{10}=10m^2+6m+1$.

 If $ x,y\in I_{1}$, then
$\frac{q+1}{2}\cdot2n<(n+2)(q+1)\leq x+yq \leq
(n+2m)(q+1)=(q+1)n+20m^2+8m<(\frac{q+1}{2}+1)\cdot2n$.
 If $ x,y\in I_{2}$, then
$(\frac{q+1}{2}+1)\cdot2n<\frac{q+1}{2}\cdot2n+20m^2+28m+8=(n+2m+2)(q+1)\leq
x+yq \leq (n+4m)(q+1)=(q+1)n+40m^2+16m<(\frac{q+1}{2}+2)\cdot2n$.
 If $ x,y\in I_{3}$, then
$(\frac{q+1}{2}+2)\cdot2n<\frac{q+1}{2}\cdot2n+40m^2+36m+8=(n+4m+2)(q+1)\leq
x+yq \leq (n+6m)(q+1)=(q+1)n+60m^2+24m<(\frac{q+1}{2}+3)\cdot2n$.

Hence, there is no skew symmetric cyclotomic cosets, and any two
cyclotomic  coset do not form a skew asymmetric pair in $T\setminus
\{C_{n}\}$. That implies that $T_{ss}=\{C_{n}\}$ and $|T_{ss}|=1$
for $0 \leq k \leq 3m$, when the defining set
$T=\bigcup_{i=0}^{k}C_{n+2i}$, where $0\leq k\leq
 3m$.\\

{\bf  Theory 3.9:}   Let  $q$ be an odd prime power of form $10m+3$,
where $m$ is a positive integer. Let $n=\frac{q^{2}+1}{10}$. Then
there exists a q-ary
$[[\frac{q^{2}+1}{10},\frac{q^{2}+1}{10}-2d+3,d;1]]$- EA-quantum MDS
codes, where $2\leq d \leq 6m+2$ is even.

 {\bf  Proof:} Consider the negacyclic codes over $F_{q^{2}}$ of length
$n=\frac{q^2+1}{10}$ with defining set
$T=\bigcup_{i=0}^{k}C_{n+2i}$, where $0 \leq k \leq 3m$ for $q$ be
an odd prime power  with the form of  $10m+3$. By Lemma 3.8, there
is $c=|T_{ss}|=1$ if $0 \leq k \leq 3m$. Since every
$q^{2}$-cyclotomic coset $C_{n+x}=\{n+x,n-x\}$, $0\leq x\leq n-1$
and $n+x$ must be odd number, we can obtain that $T$ consists of
$2k+1$ integers $\{n-2k,\cdots,n-2,n,n+2,\cdots,n+2k\}$. It implies
that $\mathcal{C}$ has minimum distance at least $2k+2$. Hence,
$\mathcal{C}$ is a $q^{2}$-ary negacyclic code with parameters
$[n,n-2(k+1)+1,\geq 2k+2]$. Combining Theory 2.7 with EA-quantum
Singleton bound, we can obtain a EA-quantum MDS code with parameters
$[[\frac{q^{2}+1}{10},\frac{q^{2}+1}{10}-2d+3,d;1]]_{q}$, where
$2\leq d \leq 6m+2$ is even.

Recently, based on constacyclic codes, zhang et al. \cite{ZhangT1}
constructed a family of new $q$-ary quantum MDS codes with
parameters $[[\frac{q^{2}+1}{5},\frac{q^{2}+1}{5}-2d+2,d]]_{q}$,
where $q=10m+3$ and $2\leq d \leq 6m+2$ is even. However, there is
no paper about the construction  of $q$-ary quantum MDS codes with
parameters $[[\frac{q^{2}+1}{10},\frac{q^{2}+1}{10}-2d+2,d]]_{q}$
with the minimal distance of  $2\leq d \leq 6m+2$ is even. From
Theory 3.8, with the help of 1 ebit, we construct a family of new
$q-$ary EA-quantum MDS codes with parameters
$[[\frac{q^{2}+1}{10},\frac{q^{2}+1}{10}-2d+3,d;1]]$, where
$q=10m+3$ and $2\leq d \leq 6m+2$.

\begin{center}
Table 4 EAQMDS codes with $n=\frac{q^{2}+1}{10}$ for $q=10m+3$ \\
\begin{tabular}{lllllllllll}
  \hline
  q                    &$[[n,k,d;1]]_{q}$                  &d                               \\
\hline
 13                    &$[[17,20-2d,d;1]]_{13}$              &$2\leq d \leq 8$ is even                 \\

 23                    &$[[53,56-2d,d;1]]_{23}$               &$2\leq d \leq 14$  is even              \\

 43                    &$[[185,188-2d,d;1]]_{43}$                &$2\leq d \leq 26$ is even            \\

 53                    &$[[281,284-2d,d;1]]_{53}$                &$2\leq d \leq 32$ is even          \\

 \hline
  \end{tabular}
\end{center}

{\bf  Lemma 3.10:}   Let $q$ be and prime power of form $10m+7$  and
$n=\frac{q^{2}+1}{10}$. If $\mathcal{C}$ is a $q^{2}$-ary negacyclic
code of length $n$ with define set $T=\bigcup_{i=0}^{k}C_{n+2i}$,
where $0\leq k\leq 3m+1$, and the decomposition of a  defining set
 $T=T_{ss}\bigcup T_{sas}$, then

\quad(i) $C_{n}$ is  skew symmetric. \quad(ii) $|T_{ss}|=1$, if $2
\leq \delta \leq 6m+4$.

{\bf  Proof:} Since $-nq\equiv
 -(\frac{q^2+1}{5}\cdot\frac{q-1}{2}-\frac{q^2+1}{10})$ $\equiv \frac{q^2+1}{10}$ $\equiv n$ mod $2n$,
  $C_{n}$ is  skew symmetric.

Let $T=\bigcup_{i=0}^{k}C_{n+2i}$, where $0\leq k\leq 3m+1$. Since
$C_{n}$ is skew symmetric, $T_{ss}$ comprises the set $\{C_{n}\}$ at
least.
 According to the concept about a  decomposition of the  defining set $T$,
 one obtain that $T_{sas}=T\backslash T_{ss}$. In order to testify $|T_{ss}| =1$ if
$0\leq k\leq 3m+1$, from Definition 2.1 and Lemma 2.2,
 we need  to testify that there is no skew symmetric cyclotomic
 coset, and any two cyclotomic
 coset do not form a  skew asymmetric pair in  $T_{sas}$.
Let $I=\{n+2i|1\leq i\leq 3m+1\}$. Only we need  to testy that for
$\forall x\in I$, $-qx$ (mod $2n$)$\not \in I$ and
$T_{ss}=\{C_{n}\}$. That implies that if $x,y\in I$, from  Lemma
2.3, $C_{x}$ is not a skew symmetric cyclotomic  coset, and any
$C_{x},C_{y}$ do not form a skew asymmetric pair if and only if
$x+yq\not\equiv0$ mod $2n$.

Divide $I$ into three parts $I_{1}=[n+2,n+2m]$,
$I_{2}=[n+2m+2,n+4m+2]$ and $I_{3}=[n+4m+4,n+6m+2]$. $q=10m+7$ and
$n=\frac{q^2+1}{10}=10m^2+14m+5$.

 If $ x,y\in I_{1}$, then
$\frac{q+1}{2}\cdot2n<(n+2)(q+1)\leq x+yq \leq
(n+2m)(q+1)=(q+1)n+20m^2+16m<(\frac{q+1}{2}+1)\cdot2n$.
 If $ x,y\in I_{2}$, then
$(\frac{q+1}{2}+1)\cdot2n<\frac{q+1}{2}\cdot2n+20m^2+36m+16=(n+2m+2)(q+1)\leq
x+yq \leq
(n+4m+2)(q+1)=(q+1)n+40m^2+52m+16<(\frac{q+1}{2}+2)\cdot2n$.
 If $ x,y\in I_{3}$, then
$(\frac{q+1}{2}+2)\cdot2n<\frac{q+1}{2}\cdot2n+2n+16m+12=(n+4m+4)(q+1)\leq
x+yq \leq (n+6m+2)(q+1)=(q+1)n+3n-16m-14<(\frac{q+1}{2}+3)\cdot2n$.

Hence, there is no skew symmetric cyclotomic cosets, and any two
cyclotomic  coset do not form a skew asymmetric pair in $T\setminus
\{C_{n}\}$. That implies that $T_{ss}=\{C_{n}\}$ and $|T_{ss}|=1$
for $0 \leq k \leq 3m+1$, when the defining set
$T=\bigcup_{i=0}^{k}C_{n+2i}$, where $0\leq k\leq
 3m+1$.\\

{\bf  Theory 3.11:}   Let  $q$ be an odd prime power of form
$10m+7$, where $m$ is a positive integer. Let
$n=\frac{q^{2}+1}{10}$. Then there exists a q-ary
$[[\frac{q^{2}+1}{10},\frac{q^{2}+1}{10}-2d+3,d;1]]$- EA-quantum MDS
codes, where $2\leq d \leq 6m+4$ is even.

 {\bf  Proof:} Consider the negacyclic codes over $F_{q^{2}}$ of length
$n=\frac{q^2+1}{10}$ with defining set
$T=\bigcup_{i=0}^{k}C_{n+2i}$, where $0 \leq k \leq 3m+1$ for $q$ be
an odd prime power  with the form of  $10m+7$. By Lemma 3.11, there
is $c=|T_{ss}|=1$ if $0 \leq k \leq 3m+1$. Since every
$q^{2}$-cyclotomic coset $C_{n+x}=\{n+x,n-x\}$, $0\leq x\leq n-1$
and $n+x$ must be odd number, we can obtain that $T$ consists of
$2k+1$ integers $\{n-2k,\cdots,n-2,n,n+2,\cdots,n+2k\}$. It implies
that $\mathcal{C}$ has minimum distance at least $2k+2$. Hence,
$\mathcal{C}$ is a $q^{2}$-ary negacyclic code with parameters
$[n,n-2(k+1)+1,\geq 2k+2]$. Combining Theory 2.7 with EA-quantum
Singleton bound, we can obtain a EA-quantum MDS code with parameters
$[[\frac{q^{2}+1}{10},\frac{q^{2}+1}{10}-2d+3,d;1]]_{q}$, where
$2\leq d \leq 6m+4$ is even.

\begin{center}
Table 5 EAQMDS codes with $n=\frac{q^{2}+1}{10}$ for $q=10m+7$ \\
\begin{tabular}{lllllllllll}
  \hline
  q                    &$[[n,k,d;1]]_{q}$                  &d                               \\
\hline
 17                    &$[[29,32-2d,d;1]]_{17}$              &$2\leq d \leq 10$ is even                 \\

 27                    &$[[73,76-2d,d;1]]_{27}$               &$2\leq d \leq 16$  is even               \\

 37                    &$[[137,140-2d,d;1]]_{37}$                &$2\leq d \leq 22$ is even            \\

 47                    &$[[221,224-2d,d;1]]_{47}$                &$2\leq d \leq 28$  is even         \\

 \hline
  \end{tabular}
\end{center}

\subsection{EA-quantum MDS Codes of Lenght
$n=\frac{q^{2}-1}{h}$ with $h\in \{3,5,7\}$}

Let $h\in \{3,5,7\}$, $q$ be an odd prime power with $h|(q+1)$.
Suppose $n=\frac{q^{2}-1}{h}$ and $r=h$. Let $\eta\in F_{q^{2}}$ be
a primitive $r^{th}$ root of unity. Since $rn=q^2-1$ clearly, every
$q^{2}$-cyclotomic coset modulo $rn$ contains exactly one element.
In \cite{Chen}, using Hermitian construction and $\eta$-constacyclic
 codes, Chen et al. have constructed a
family of new quantum MDS codes with parameters
$[[\frac{q^{2}-1}{h},\frac{q^{2}-1}{h}-2d+2,d]]_{q}$ , where $q$ is
an odd prime power with $h|(q+1)$, $h\in \{3,5,7\}$ and $2\leq d\leq
\frac{(q+1)(h+1)}{2h}-1$.  It is very hard to enlarge the minimal
distance for this type quantum MDS codes. In this subsection, adding
one ebit, we construct a new family of a family of new EA-quantum
MDS codes with parameters
$[[\frac{q^{2}-1}{h},\frac{q^{2}-1}{h}-2d+3,d;1]]_{q}$, where
$\frac{(q+1)(h-1)}{2h}+1\leq d \leq \frac{(q+1)(h+3)}{2h}-1$.

{\bf  Lemma 3.12:}   Let $q$ is an odd prime power with $h|(q+1)$,
$h\in \{3,5,7\}$ and $n=\frac{q^{2}-1}{h}$. If $\mathcal{C}$ is a
$q^{2}$-ary constacyclic code of length $n$ with define set
$T=\bigcup_{i=\frac{(h-3)(q+1)}{2h}}^{k}\{C_{1+hi}\}$, where
$\frac{(h-3)(q+1)}{2h}\leq k \leq q-2$, and the decomposition of a
defining set
 $T=T_{ss}\bigcup T_{sas}$, then

\quad(i) $C_{1+(\frac{(h-1)(q+1)}{2h}-1)h}$ is skew symmetric.
\quad(ii) $|T_{ss}|=1$, if $\frac{(h-1)(q+1)}{2h}-1\leq k \leq q-2$.

{\bf  Proof:} (i) Since
$1+(\frac{(h-1)(q+1)}{2h}-1)h=\frac{(h-1)(q-1)}{2}$ and
$-[1+(\frac{(h-1)(q+1)}{2h}-1)h]q\equiv
 -\frac{(h-1)(q-1)}{2}q$ $\equiv -[\frac{h-1}{2}(q^{2}-1)-\frac{(h-1)(q-1)}{2}]$
  $\equiv \frac{(h-1)(q-1)}{2}$  $\equiv 1+(\frac{(h-1)(q+1)}{2h}-1)h$ mod $hn$,
  $C_{1+(\frac{(h-1)(q+1)}{2h}-1)h}$ is  skew symmetric.

(ii) Let $T=\bigcup_{i=\frac{(h-3)(q+1)}{2h}}^{k}\{C_{1+hi}\}$,
where $\frac{(h-3)(q+1)}{2h}\leq k \leq q-2$. Since
$C_{1+(\frac{(h-1)(q+1)}{2h}-1)h}$ is skew symmetric, $T_{ss}$ comprises the set $\{C_{1+(\frac{(h-1)(q+1)}{2h}-1)h}\}$ at
least.
 According to the concept about a  decomposition of the  defining set $T$,
 one obtain that $T_{sas}=T\backslash T_{ss}$.
 In order to testify $|T_{ss}| =1$ if $\frac{(h-1)(q+1)}{2h}-1\leq i \leq q-2$, from Definition 2.1 and Lemma 2.2,
 we need  to testify that there is no skew symmetric cyclotomic
 coset, and any two cyclotomic
 coset do not form a  skew asymmetric pair in  $T_{sas}$.

Let $I=\{1+hi| \frac{(h-3)(q+1)}{2h}\leq i \leq q-2\}\setminus (1+(\frac{(h-1)(q+1)}{2h}-1)h)$ and $r=h$.
Only we need  to testy that for
$\forall x\in I$, $-qx$ (mod $rn$)$\not \in I$ and
$T_{ss}=\{C_{1+(\frac{(h-1)(q+1)}{2h}-1)h}\}$. That implies that if $x,y\in I$, from  Lemma
2.3, $C_{x}$ is not a skew symmetric cyclotomic  coset, and any
$C_{x},C_{y}$ do not form a skew asymmetric pair if and only if
$x+yq\not\equiv0$ mod $2n$.

Divide $I$ into three parts $I_{1}=[1+\frac{(h-3)(q+1)}{2h}h,1+(\frac{(h-1)(q+1)}{2h}-2)h]$,
$I_{2}=[1+(\frac{(h-1)(q+1)}{2h})h,1+(\frac{(h+1)(q+1)}{2h}-2)h]$ and $I_{3}=[1+(\frac{(h+1)(q+1)}{2h}-1)h,1+(q-2)h]$.
$rn=hn=q^2-1$.
 If $ x,y\in I_{1}$, then
$\frac{(h-3)}{2}(q^{2}-1)+(h-2)(q+1)=(1+\frac{(h-3)(q+1)}{2h}h)(q+1)\leq x+yq \leq
(1+(\frac{(h-1)(q+1)}{2h}-2)h)(q+1)=\frac{(h-1)}{2}(q^{2}-1)-h(q+1)$.
Since $h\in\{3,5,7\}$,  $\frac{(h-3)}{2}=\{0,1,2\}$, and $\frac{(h-1)}{2}=\{1,2,3\}$.
Hence, for $ x,y\in I_{1}$,  $ln< x+yq < (l+1)n$, where $0\leq l\leq2$ is a integer.

 If $ x,y\in I_{2}$, then
$\frac{h-1}{2}(q^{2}-1)+h(q+1)=(1+(\frac{(h-1)(q+1)}{2h})h)(q+1)\leq
x+yq \leq (1+(\frac{(h+1)(q+1)}{2h}-2)h)(q+1)=\frac{h+1}{2}(q^{2}-1)-(h-2)(q+1)$.
Since $h\in\{3,5,7\}$,  $\frac{(h-1)}{2}=\{1,2,3\}$, and $\frac{(h+1)}{2}=\{2,3,4\}$.
Hence, for $ x,y\in I_{2}$,  $ln< x+yq < (l+1)n$, where $1\leq l\leq 3$ is a integer.

For $h=3$, if $ x,y\in I_{3}$, then
$2n<\frac{h+1}{2}(q^{2}-1)+2(q+1)=(1+(\frac{(h+1)(q+1)}{2h}-1)h)(q+1)\leq
x+yq \leq (1+(q-2)h)(q+1)=h(q^{2}-1)-(h-1)(q+1)<3n$.

For $h=5$,  Divide $I_{3}$ into two parts $I_{3}^{'}=[1+(\frac{(h+1)(q+1)}{2h}-1)h,1+(\frac{(h+3)(q+1)}{2h}-2)h]$ and
$I_{3}^{''}=[1+(\frac{(h+3)(q+1)}{2h}-1)h,1+(q-2)h]$.

if $ x,y\in I_{3}^{'}$, then
$3n<\frac{h+1}{2}(q^{2}-1)+2(q+1)=(1+(\frac{(h+1)(q+1)}{2h}-1)h)(q+1)\leq
x+yq \leq (1+(\frac{(h+3)(q+1)}{2h}-2)h)(q+1)=\frac{h+3}{2}(q^{2}-1)-(h-2)(q+1)<4n$;
if $ x,y\in I_{3}^{''}$,
$4n<(1+(\frac{(h+3)(q+1)}{2h}-2)h)(q+1)\leq
x+yq \leq (1+(q-2)h)(q+1)=h(q^{2}-1)-(h-1)(q+1)<5n$.

For $h=7$,  Divide $I_{3}$ into two parts $I_{3}^{a}=[1+(\frac{(h+1)(q+1)}{2h}-1)h,1+(\frac{(h+3)(q+1)}{2h}-2)h]$,
$I_{3}^{b}=[1+(\frac{(h+3)(q+1)}{2h}-1)h,1+(\frac{(h+5)(q+1)}{2h}-2)h]$ and $I_{3}^{c}=[1+(\frac{(h+5)(q+1)}{2h}-1)h,1+(q-2)h]$.

If $ x,y\in I_{3}^{a}$, $4n<\frac{h+1}{2}(q^{2}-1)+2(q+1)=(1+(\frac{(h+1)(q+1)}{2h}-1)h)h)(q+1)\leq
x+yq \leq (1+(\frac{(h+3)(q+1)}{2h}-2)h)(q+1)=\frac{h+3}{2}(q^{2}-1)-(h-2)(q+1)<5n$;
if $ x,y\in I_{3}^{b}$,
$5n<(1+(\frac{(h+3)(q+1)}{2h}-2)h)(q+1)\leq
x+yq \leq (1+(\frac{(h+5)(q+1)}{2h}-2)h)(q+1)=\frac{h+5}{2}(q^{2}-1)-(h-2)(q+1)<6n$.
if $ x,y\in I_{3}^{c}$,
$6n<(1+(\frac{(h+3)(q+1)}{2h}-1)h)(q+1)\leq
x+yq \leq (1+(q-2)h)(q+1)=h(q^{2}-1)-(h-1)(q+1)<7n$.

Hence, there is no skew symmetric cyclotomic cosets, and any two
cyclotomic  coset do not form a skew asymmetric pair in $T\setminus
\{C_{1+(\frac{(h-1)(q+1)}{2h}-1)h}\}$. That implies that $T_{ss}=\{C_{1+(\frac{(h-1)(q+1)}{2h}-1)h}\}$ and $|T_{ss}|=1$
for $\frac{(h-1)(q+1)}{2h}-1\leq i \leq q-2$, when the defining set
$T=\bigcup_{i=\frac{(h-3)(q+1)}{2h}}^{k}\{C_{1+hi}\}$,
where $\frac{(h-3)(q+1)}{2h}\leq k \leq q-2$.\\

{\bf  Theory 3.13:}  Let $q$ is an odd prime power with $h|(q+1)$,
$h\in \{3,5,7\}$ and $n=\frac{q^{2}-1}{h}$.
 Then there exists a q-ary
$[[\frac{q^{2}-1}{h},\frac{q^{2}-1}{h}-2d+3,d;1]]$, where
$\frac{q+1}{h}+1\leq d \leq \frac{(h+3)(q+1)}{2h}-1$.

{\bf  Proof:} Consider the constacyclic codes over $F_{q^{2}}$ of
length $n=\frac{q^2-1}{h}$ with defining set
$T=\bigcup_{i=\frac{(h-3)(q+1)}{2h}}^{k}\{C_{1+hi}\}$, where
$\frac{(h-3)(q+1)}{2h}\leq k \leq q-2$, $h\in \{3,5,7\}$ and
$h|(q+1)$ for $q$ be an odd prime power. By Lemma 3.12, there is
$c=|T_{ss}|=1$ if $\frac{(h-1)(q+1)}{2h}-1\leq k \leq q-2$. Since
every $q^{2}$-cyclotomic coset has one element which must be odd
number, we can obtain that $T$ consists of
$(i-\frac{(h-3)(q+1)}{2h}+1)$ integers
$\{1+(\frac{(h-3)(q+1)}{2h})h,1+(\frac{(h-3)(q+1)}{2h}+1)h,
1+(\frac{(h-3)(q+1)}{2h}+2)h\cdots,1+kh\}$. It implies that
$\mathcal{C}$ has minimum distance at least
$(i-\frac{(h-3)(q+1)}{2h}+2)$. Hence, $\mathcal{C}$ is a $q^{2}$-ary
constacyclic code with parameters
$[n,n-2(i-\frac{(h-3)(q+1)}{2h}+1)+1,\geq
(i-\frac{(h-3)(q+1)}{2h}+2)]$. Combining Theory 2.7 with EA-quantum
Singleton bound, we can obtain a EA-quantum MDS code with parameters
$[[\frac{q^{2}-1}{h},\frac{q^{2}-1}{h}-2d+3,d;1]]_{q}$, where
$\frac{q+1}{h}+1\leq d \leq \frac{(h+3)(q+1)}{2h}-1$.\\

\begin{center}
Table 6 EAQMDS codes with $n=\frac{q^{2}-1}{h}$ \\
\begin{tabular}{lllllllllll}
  \hline
  q          & h          &$[[n,k,d;1]]_{q}$                  &d                               \\
\hline
 11          & 3          &$[[40,43-2d,d;1]]_{11}$              &$5\leq d \leq 11$                  \\

 17          & 3          &$[[96,99-2d,d;1]]_{17}$               &$7\leq d \leq 17$               \\

 19           &5          &$[[72,75-2d,d;1]]_{19}$                &$5\leq d \leq 15$             \\

 29           &5          &$[[168,171-2d,d;1]]_{29}$                &$7\leq d \leq 23$         \\

 13           &7          &$[[24,27-2d,d;1]]_{13}$                &$3\leq d \leq 9$             \\

 41           &7          &$[[240,243-2d,d;1]]_{41}$                &$7\leq d \leq 29$         \\

 \hline
  \end{tabular}
\end{center}

\section{ SUMMARY}

In this paper, we propose a concept about a decomposition of the
defining set of constacyclic codes. Using this concept, we construct
four classes of $q$-ary entanglement-assisted quantum MDS (EAQMDS)
codes based on classical constacyclic MDS codes by exploiting less
pre-shared maximally entangled states.

In Table 7, we list the $q$-ary entanglement-assisted quantum MDS
codes constructed in this paper. By consuming four pre-shared
maximally entangled states, we obtain the EA-quantum MDS codes of
$n=q^2+1$ with the minimal distance upper limit greater than  $3q-1$
(even). These EA-quantum MDS codes are improved the parameters of
codes in Ref.\cite{Chen2}. Consumed only one pair of maximally
entangled states,  each EA-quantum MDS code of length
$n=\frac{q^2+1}{10}$ has half length comparing with  the standard
QMDS code of the same minimum distance constructed in
Ref.\cite{Chen}. Moreover, consuming only one pair of maximally
entangled states, we obtain a family of EA-quantum MDS codes of
$\frac{q^2-1}{h}$ with the minimal distance larger than the standard
quantum MDS codes in Ref.\cite{Chen}.

Comparing the
parameters with all known $q$-ary EA-quantum MDS codes, we find that these
quantum MDS  codes are new in the sense that their parameters are
not covered by the codes available in the literature.\\

\begin{center}
Table 7 New parameters of EAQMDS codes \\
\begin{tabular}{lllllllllllllll}
  \hline
  Class                    & length                      &$[[n,k,d;c]]_{q}$                &Distance  \\
\hline
  1                        & $q^2+1$                     &$[[q^2+1,q^2-2d+7,d;4]]_{q}$      &$q+3\leq d \leq 3q-1$ is even \\
                           &               &               &         \\

  2                        &$\frac{q^2+1}{10}$,$q=10m+3$  &$[[\frac{q^2+1}{10},\frac{q^2+1}{10}-2d+3,d;1]]_{q}$  &$2\leq d \leq 4m+2$ is even\\
                           &               &               &         \\

 3                        &$\frac{q^2+1}{10}$,$q=10m+7$    &$[[\frac{q^2+1}{10},\frac{q^2+1}{10}-2d+3,d;1]]_{q}$  &$2\leq d \leq 4m+4$ is even\\
                           &               &   \\

 4                        &$n=\frac{q^{2}-1}{h}$        &$[[\frac{q^2-1}{h},\frac{q^2-1}{h}-2d+3,d;1]]_{q}$    &$\frac{q+1}{h}+1\leq d \leq \frac{(h+3)(q+1)}{2h}-1$\\
                          &$h\in \{3,5,7\}$   & &                \\

  \hline
  \end{tabular}
\end{center}

{\bf Remark:} This paper was submitted to the Journal "Designs,
Codes and Cryptography" on Jan. 11, 2018.

\section*{Acknowledgment}
This work is supported by the National Key R\&D Program of China
under Grant No. 2017YFB0802400, the National Natural Science
Foundation of China under grant No. 61373171, the National Natural
Science Foundation of China under Grant No.11471011, the Natural
Science Foundation of Shaanxi under Grant No.2017JQ1032, 111
Project under grant No.B08038, and the
Key Project of Science Research of Anhui Province(Grant No.KJ2017A519).

\bibliographystyle{amsplain}

\end{document}